\documentclass[journal]{new-aiaa}
\usepackage[utf8]{inputenc}
\usepackage{textcomp}

\usepackage{graphicx}
\usepackage{amsmath}
\usepackage[version=4]{mhchem}
\usepackage{siunitx}
\usepackage{longtable,tabularx}
\title{Long-Term Earth Magnetosphere Science Orbit via Earth-Moon Resonance Orbit}

\author{Jinsung Lee \footnote{Research Scientist, Satellite Technology Research Center, jinsung\_lee@kaist.ac.kr, AIAA Member, Corresponding Author}}
\affil{Korea Advanced Institute of Science and Technology, Daejeon, 34141, South Korea}

\author{Jaeyoung Kwak\footnote{Ph.D. Candidate, University of Science and Technology, 776 Daedeok-daero}}
\affil{University of Science and Technology, Daejeon 34055, South Korea}

\author{Jaemyung Ahn\footnote{Professor, Department of Aerospace Engineering, jaemyung.ahn@kaist.ac.kr, AIAA Associate Fellow}}
\affil{Korea Advanced Institute of Science and Technology, Daejeon, 34141, South Korea}

\begin{document}

\maketitle

\begin{abstract}
This article investigates long-term orbits within the Earth's magnetosphere, specifically focusing on orbits where the argument of periapsis is synchronized with changes induced by lunar gravity assists and the Earth's argument of latitude over a complete orbital period in Earth-Moon resonance.
In the Earth-Moon rotating frame, resonance orbits appear repetitive; however, the argument of periapsis shifts due to the third-body effects from lunar flybys. The extent of this shift is influenced by the Jacobi integral associated with the resonance orbit.
To identify feasible resonance orbits and the optimal Jacobi integral, we map the argument of periapsis change against the Jacobi integral for each prospective orbit. This synchronization allows the spacecraft to remain within a confined region in space when observed from the Sun-Earth rotating frame.
Finally, the article discusses the applications of these long-term Earth magnetosphere science orbits, including orbit-orientation reconfiguration (station keeping) and stability.
\end{abstract}

\section*{Nomenclature}

{\renewcommand\arraystretch{1.0}
\noindent\begin{longtable*}{@{}l @{\quad=\quad} l@{}}
$C$ & Jacobi integral \\ 
$SEHO$ & Sun-Earth Harmonic Orbit\\
$SSO$ & Sun-Synchronous Orbit\\
$CRTBP$ & Circular-Restricted Three-Body Problem\\
$\Omega$ & Potential function of CRTBP\\
$C$ & Jacobi integral\\
$n_e$ & Mean orbital motion of the Earth\\
$N$ & Number of Moon's orbit\\
$M$ & Number of spacecraft's orbit\\
$N:M$ & Resonance orbit classifier\\
$\Delta\omega$ & Change in the argument of periapsis\\
$T$ & Orbital period
\end{longtable*}}

\section{Introduction}
\label{sec1}
We present a new orbit class called the Sun-Earth Harmonic Orbit (SEHO) \cite{lee2024sun}. The SEHO utilizes Earth-Moon resonance orbits to constrain its orbital motion within a specific region in the Sun-Earth rotating frame. Similar to the Sun-Synchronous Orbit (SSO), the SEHO maintains a consistent position relative to the Sun throughout the year. However, SEHO and SSO differ significantly in their orbital shape and precession mechanisms.

The SSO leverages J2 and J4 perturbations to match the instantaneous precession rate of the right ascension of the ascending node with the Earth's mean orbital motion around the Sun. Typically, SSOs are executed with circular orbits and carefully chosen inclinations, although theoretically, they can be eccentric. Using the equation for the precession rate of the right ascension of the ascending node \cite{vallado2001fundamentals}, the maximum allowable semi-major axis for a circular SSO is approximately 12,352 km, extending to about 15,400 km for eccentric orbits (e $\approx$ 0.55). Beyond these limits, the orbit becomes prograde (inclination $< 180^\circ$ ), negating the advantages of SSO. Moreover, such eccentric and high-altitude orbits are unlikely to exhibit SSO characteristics due to third-body effects and gravitational perturbations from the Moon and Sun.

In contrast, SEHO employs Earth-Moon resonance orbits to emulate SSO-like motion for highly eccentric orbits with large semi-major axes. While the resonance orbit repeats when viewed from the Earth-Moon rotating frame, it experiences changes in the argument of periapsis in the Earth-centered inertial (ECI) frame due to consistent lunar flybys. The magnitude of this rotation depends on the Jacobi Integral and the flyby altitude of the resonance orbit. SEHO compensates for the Earth's true anomaly change during one orbital period of the resonance orbit by utilizing this argument of periapsis rotation caused by lunar flybys. Unlike SSO, which alters the right ascension of the ascending node, SEHO modifies the argument of periapsis. Consequently, when viewed from the Sun, the eccentric orbit is not stationary but oscillates by following the Moon's orbital plane with respect to the Sun.

Highly eccentric orbits are particularly well-suited for Earth's magnetospheric science missions. Numerous space missions have aimed to explore the regions where Earth's magnetic field interacts with the solar wind and Coronal Mass Ejections (CMEs), such as the magnetopause and bow shock layer. Previous missions, including Magnetospheric Multiscale (MMS) \cite{fuselier2016magnetospheric}\cite{burch2016magnetospheric}, Cluster II \cite{escoubet2001introduction}, and Time History of Events and Macroscale Interactions during Substorms (THEMIS) \cite{angelopoulos2008first}\cite{sibeck2008themis}, have utilized highly eccentric orbits to collect scientific measurements in these regions. However, unlike SSO, these science orbits remain stationary in the inertial frame without precession, limiting spacecraft transit through the most scientifically relevant regions to only half or a quarter of the year.

For instance, MMS's primary objective was to measure magnetic reconnection within the magnetosphere \cite{sibeck2008themis}, which occurs on the far side of the magnetopause from the Sun. Unfortunately, the orbit design allowed the spacecraft to pass through the most prominent region of Earth's magnetosphere for its primary science objective during only a quarter of the year. SEHO aims to address this limitation by modifying the argument of periapsis of highly eccentric orbits through multiple lunar flybys. This approach maintains the science orbit's orientation consistent with the bow shock and magnetopause orientation, enabling spacecraft to collect measurements within the most prominent magnetopause region throughout the year.

This paper is structured as follows: Section 2 further discusses the scientific advancements achievable by placing spacecraft (or a swarm of spacecraft) in the proposed orbit. Section 3 introduces the dynamical and coordinate systems of the Circular-Restricted Three-Body Problem and resonance orbits.
Section 4 examines the graphical representation of the argument of periapsis rotation and the mean motion of the Earth, which is used to determine the optimal Jacobi integral for determining the optimal Sun-Earth Harmonic Orbit.
We then present three SEHOs with Earth-Moon resonance orbits of N:M=1:2, 3:7, and 2:5 along with each orbit's characteristics. Finally, Section 5 examines the stability of SEHO using Broucke's stability index \cite{broucke1969stability}, calculated via the Monodromy matrix of the optimal orbit. This section also explores practical station-keeping strategies for maintaining the optimal resonance orbit, focusing on leveraging lunar swingby maneuvers.

\section{Scientific Advances of Sun-Earth Harmonic Orbit} 
Several planets orbiting the Sun possess magnetic shields, known as magnetospheres, which protect them from electrically charged particles such as galactic cosmic rays (GCR) and solar energetic particles (SEP). Without these magnetospheres, high-energy charged particles would bombard planetary surfaces, potentially stripping away neutral particles and rendering life impossible by acting as a radiation source. Earth's magnetosphere, shaped by the solar wind, forms a highly asymmetric ellipsoid extending approximately 10 Earth radii ($R_E$) in the sunward direction and over 100 $R_E$ in the anti-sunward direction under normal conditions.
The magnetopause, the boundary between the interplanetary magnetic field (IMF) and Earth's magnetosphere, forms when Earth's magnetic field deflects charged particles comprising the solar wind. 

At the magnetopause, the different gyration directions of electrons and protons create a current known as the magnetopause current. When the IMF is directed southward, its magnetic field lines oppose Earth's magnetic field, leading to a significant reduction in magnetosphere size—potentially shrinking it to geosynchronous orbit—due to magnetic field reconnection at the magnetopause \cite{kim2024localized}. This reconnection phenomenon typically develops complicated micro-scale magnetic flux ropes, allowing accelerated solar wind to penetrate the magnetosphere \cite{dokgo2021waves}. Even with a northward IMF, an instability known as the Kelvin-Helmholtz instability (KHI) can occur at the magnetosphere's flanks due to flow velocity differences on either side of the boundary. As reported by \cite{hasegawa2009kelvin} and \cite{hwang2012first}, KHI-driven rolled-up vortices or waves (i.e., Kelvin-Helmholtz waves; KHW) can be generated at the magnetopause. These plasma processes occur within the magnetosheath, the transition region between the bow shock and the magnetopause.
The magnetosphere's flanks extend into the anti-sunward region known as the magnetotail. Charged particles conveyed by magnetic field reconnection at the magnetopause accumulate within the magnetotail, forming a dawn-to-dusk electric field. Particle flow along this electric field is known as the magnetotail current, and these particles subsequently drift toward Earth. The charged particles in the magnetotail region are primarily accelerated by magnetic reconnection mechanisms at approximately 100 $R_E$. During geomagnetic substorm periods, these particles are further accelerated by magnetic field reconnection at the near-Earth neutral line (NENL) around 10 $R_E$ \cite{baker1996neutral} before precipitating into the sub-polar ionosphere region, generating auroras. Consequently, magnetotail studies can be conducted in conjunction with auroral process investigations using data from ground-based all-sky imagers \cite{miyashita2009state, talha2024association}.

Regarding in-situ research, past and current operational satellites have typically focused on specific research topics. For example, the Cluster mission concentrated on plasma dynamics in the polar regions of the magnetopause and cusp, where solar wind can freely penetrate Earth's high latitudes, using a relatively high-inclination orbit \cite{https://doi.org/10.1029/2021JA029362}. The MMS mission successfully observed both magnetopause and magnetotail regions, albeit at different times. Similarly, the THEMIS mission aimed to understand processes occurring during magnetic field reconnection in the magnetotail region.
In-situ observation using SEHO offers two main advantages compared to other missions with similar scientific objectives. First, unlike other missions, SEHO passes through nearly the same location in the Sun-Earth rotating frame with each orbit, enabling periodic area monitoring. Second, SEHO is unique in its ability to observe both magnetopause and magnetotail regions within the same orbit, making it ideal for studying overall magnetospheric changes during geomagnetic storm periods. As mentioned earlier, the spatial extent of the magnetosphere is highly variable, depending on solar activity. These dynamic changes typically unfold over a few days, which aligns with SEHO's orbital period. Therefore, using a swarm of satellites in SEHO (for example, three satellites) simultaneously makes it possible to compare and monitor changes on the dawn and dusk flanks of the magnetopause and magnetotail during the same magnetospheric events.

Recently, in addition to in-situ observations, remote observations have been planned to provide a global view of the changing magnetosphere. For example, the Lunar Environment and Heliospheric X-ray Imager (LEXI), a payload on a future lunar surface exploration mission, will observe soft X-rays (0.1-2 keV) produced by charge-exchange between solar winds and exospheric neutrons to monitor the dayside magnetopause \cite{walsh2024lunar}. The Solar Terrestrial Observer for the Response of the Magnetosphere (STORM) mission has broader objectives, aiming to observe the magnetopause, magnetotail, polar region aurora, and inner plasmasphere using Far Ultraviolet (FUV) spectrographic telescopes, X-ray Imagers (XRI), Energetic Neutral Atom (ENA) cameras, and Extreme Ultraviolet (EUV) imagers in a nearly 90-degree inclination polar orbit \cite{sibeck2023quantifying}.

In this context, science missions using SEHO can provide excellent remote observation of the dayside magnetopause region through X-ray monitoring. Unlike observations from the lunar surface or lunar orbit, SEHO enables satellites to monitor target regions relatively frequently. By operating multiple satellites simultaneously, continuous and uninterrupted observations become possible. Combined with in-situ observations, this multi-satellite SEHO approach would effectively enhance our understanding and ability to compare overall reactions in different parts of the magnetosphere for various types of magnetospheric changes.

\section{Dynamical systems of Circular Restricted Three-Body Problem}
The SEHO orbits are designed based on the Earth-Moon resonance orbit, which exists under the circular restricted three-body problem (CRTBP). The CRTBP studies the satellite's motion under the gravitational influence of the two large bodies (i.e., Sun-Earth or Earth-Moon). The CRTBP coordinate frame is a rotating frame. The x-axis is defined by the direction vector from the primary to the secondary body, the z-axis by the secondary body’s angular momentum, and the y-axis from the cross-product of the two (z and x) axes. The governing equations of the CRTBP dynamics are as follows.
\begin{equation}
{\bf{\dot x}} = f\left( {\bf{x}} \right) = {\left[ {\begin{array}{*{20}{c}}
{{\bf{v}},}&{2{v_y} + \frac{{\partial {\Omega _3}}}{{\partial x}},}&{ - 2{v_x} + \frac{{\partial {\Omega _3}}}{{\partial y}},}&{\frac{{\partial {\Omega _3}}}{{\partial z}}}
\end{array}} \right]^T}
\end{equation}
The state vector ${\bf{x}}\left( { = {{\left[ {x,y,z,{v_x},{v_y},{v_z}} \right]}^T}} \right)$ represents the position and velocity components in the rotating frame. The potential function of the CRTBP ($\Omega_3$) takes the following form
\begin{equation}
    {\Omega _3}\left( {x,y,z} \right) = \frac{{{x^2} + {y^2}}}{2} + \frac{{1 - \mu }}{{{r_1}}} + \frac{\mu }{{{r_2}}}
    \label{eq:omega3}
\end{equation}
The mass ratio ($\mu \left( { = {m_2}/({m_1} + {m_2})} \right)$) is defined by the mass ratio of the secondary body over the sum of two large masses where the primary is considered heavier ($m_1 > m_2$). ${r_1}$ and ${r_2}$ measures the normalized distance from the primary and secondary masses to the spacecraft as follows.

\begin{equation}
    {r_1} { = \sqrt {{{\left( {x + \mu } \right)}^2} + {y^2} + {z^2}} } 
\end{equation}
\begin{equation}
    {r_2} { = \sqrt {{{\left( {x - 1 + \mu } \right)}^2} + {y^2} + {z^2}} } 
\end{equation}

A resonance orbit is a unique coupling orbit where the spacecraft and the secondary body orbit around their central planet by a small integer ratio of N (number of Moon’s revolutions) and M (number of spacecraft’s revolutions). The secondary body perfectly pushes and pulls the spacecraft during the spacecraft’s closest approach to the secondary body, resulting in a repeating trajectory when viewed from the rotating reference frame. The orbital period of the resonance orbit, which depends on the resonance orbit ratio (N:M) and the Jacobi integral, is defined by time duration until the trajectory returns to its original location in the rotating frame. This repeating trajectory has a symmetric shape about the x-axis of the rotating frame. An infinite number of resonance orbit ratios exist, and an infinite amount of single resonance orbit ratios exist with varying Jacobi integral. Jacobi’s integral is a conserved quantity used to define the Hill’s region, which is used to visualize the forbidden and allowed regions in the rotation frame. The Jacobi integral is defined as follows.
\begin{equation}
C = 2{\Omega _3} - (v_x^2 + v_y^2 + v_z^2)
\end{equation}
Resonance orbit with a smaller Jacobi integral transits closer to the secondary body than the larger Jacobi integral. There are two resonance orbits: the interior resonance orbit ($N<M$) and the exterior resonance orbit($N>M$). In this study, the interior resonance orbit is mainly discussed.

\begin{figure}
    \centering
    \includegraphics[page=1, width=1\linewidth]{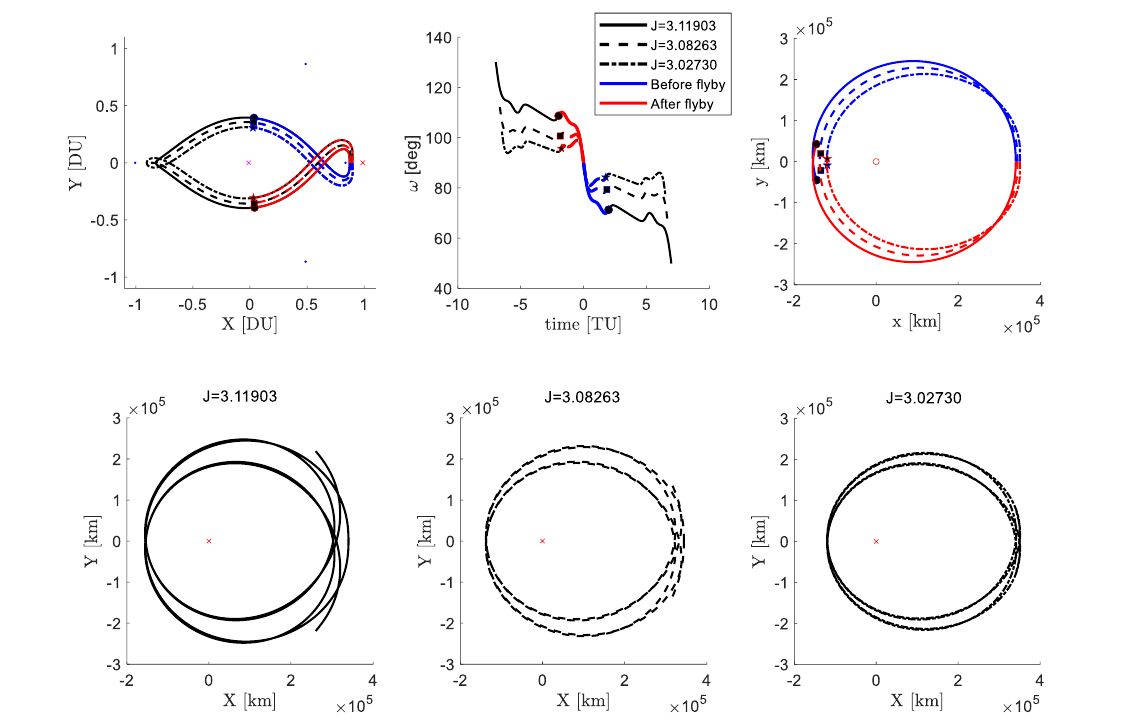}
    \caption{Left top: N:M=1:2 Earth-Moon resonance orbit in Earth-Moon rotating frame with different Jacobi integral. Middle top: argument of periapsis changes in between two periapsis before and after the lunar flyby. Right top: N:M=1:2 trajectory in ECI frame, propagation halted at periapsis. Bottom: N:M=1:2 resonance orbit with different Jacobi integral plotted for two $T(N:M,C)$ in ECI frame. A larger change in the argument of periapsis can be seen with a smaller Jacobi integral}
    \label{fig:1}
\end{figure}

In our intuition, spacecraft that travels closer to the secondary body would experience more significant prior and posterior changes in their orbital elements. However, such intuition is not accurate for all orbital elements. Figure \ref{fig:1} shows the change in the argument of periapsis ($\omega$) as a function of time and its trajectory plotted in the rotating frame with different Jacobi integrals. Note that the trajectory with a larger Jacobi integral, which travels further away during the flyby, experiences a larger change in the argument of periapsis. This intuition error occurs because the trajectories are viewed in the rotating frame. The change in the argument of periapsis ($\Delta\omega$) for each resonance orbit was determined by performing forward and backward propagation from the initial condition of the resonance orbit, which is on the positive x-axis ($y=0,\dot x =0$). The propagation was performed until the trajectory resembled a periapsis under the Keplerian orbit assumption (plotted in red (backward) and blue (forward) lines in figure \ref{fig:1}). $\Delta\omega$ was calculated for all resonance orbits of interest, which were 1:2, 3:7, 2:5, 3:8 and 1:3 with varying Jacobi’s integral. As the orbital period ($T$) and argument of periapsis change varies between resonance orbits and Jacobi’s integral, the following term is used throughout this paper: $T(N:M,C)$ and $\Delta\omega(N:M,C)$.

\section{Characteristics of Sun-Earth Harmonic Orbit}

\begin{figure}
    \centering
    \includegraphics[page=2, width=1\linewidth]{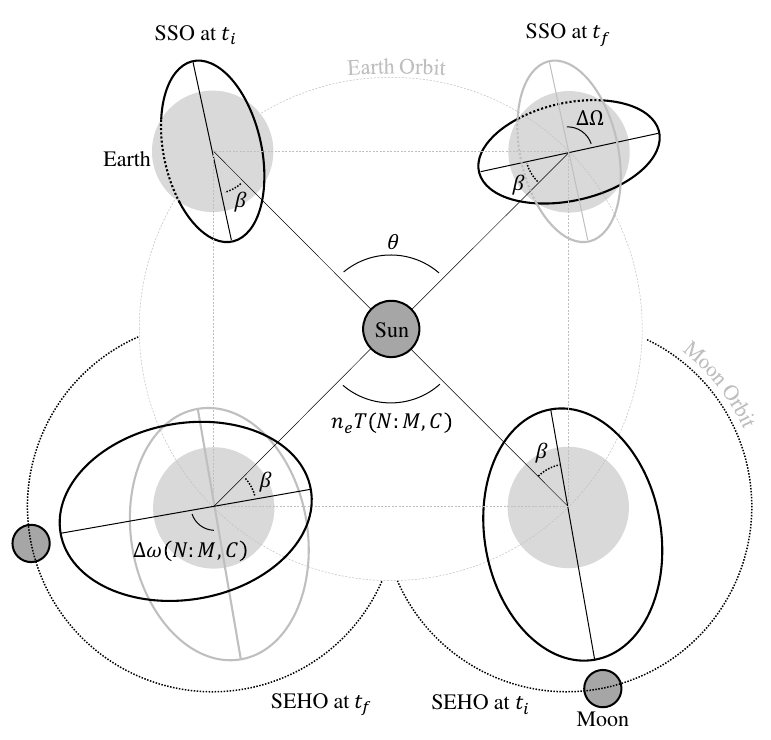}
    \caption{Comparison of SEHO and SSO in SCI frame}
    \label{fig:comparison}
\end{figure}

The SEHO corresponds to an Earth-moon resonance orbit which generates $\Delta\omega(N:M,C)$ identical to $n_e T(N:M,C)$ during one orbital period of the Earth-Moon resonance orbit (see figure \ref{fig:comparison}), where $n_e$ and $T(N:M,C)$ denotes the mean orbital motion of the Earth around the Sun and the orbital period of the resonance orbit in the rotating frame, respectively. For SSOs, the Sun angle, $\beta$, is the angular offset between the Sun vector projected to the orbital plane and the Sun vector itself. When the right ascension ascending node’s precession rate is matched with the natural orbital motion of the Earth around the Sun, the Sun angle is held constant. However, the SEHO’s sun angle, measured from the angular difference between the eccentricity vector of the resonance orbit and the Sun vector, is not constant but compensated during each lunar flyby.

\begin{figure}
    \centering
    \includegraphics[page=3, width=1\linewidth]{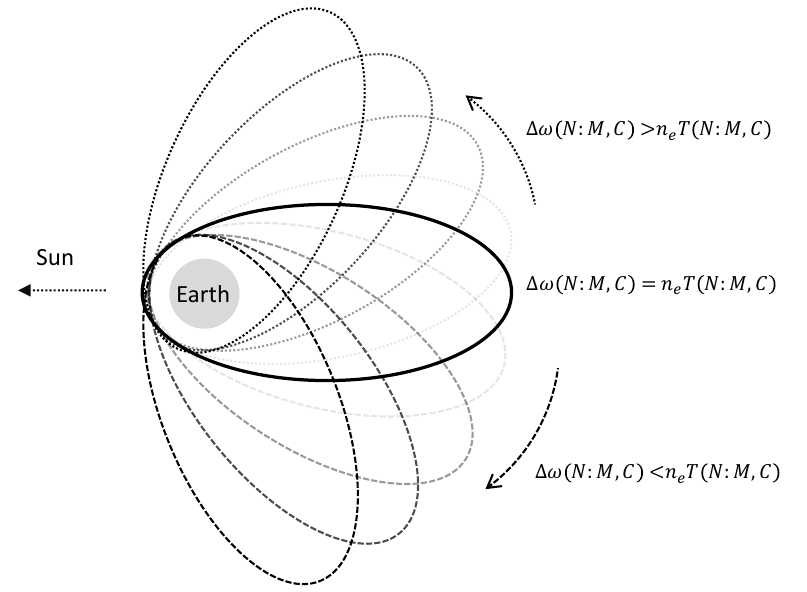}
    \caption{Trajectory shifting trend based on argument of periapsis change}
    \label{fig:rotation}
\end{figure}

Unlike the SSO, the SEHO’s orbital plane is stationary to the Moon’s orbital plane and inertial to the Sun's reference frame (SCI). Therefore, the SEHO orbits are always in inclination of 5.34 degrees, which is the orbital inclination of the Moon measured from the ecliptic plane. When $n_e T(N:M,C)$ and $\Delta\omega(N:M,C)$ from lunar flyby are not equal, the orbit experiences orbital orientation changes when viewed from the Sun-Earth rotating frame (see Figure \ref{fig:rotation}). The orientation rotates clockwise when $\Delta \omega \left( {N:M,C} \right) < {n_e}T\left( {N:M,C} \right)$ and counter-clock-wise when $\Delta \omega \left( {N:M,C} \right) > {n_e}T\left( {N:M,C} \right)$. Such orientation shift can be used to our advantage for performing station-keeping maneuvers or changing the orientation for performing science in differently oriented SEHO, which is discussed in a later chapter.

The orientation of the SEHO can be chosen depending on the designated mission type. For space telescope and Earth magnetosphere missions, where the missions benefit by having the apoapsis always point away from the Sun, one could choose a lunch date that places the lunar gravity assist aligned with the Sun-Earth vector with the Moon in the opposite direction of the Sun. SEHO’s trajectory that utilizes interior Earth-Moon resonance orbits travels within the Earth magnetopause region and stays above the Van Allen radiation belt. Similarly, missions such as Solar observatories could benefit by having the apoapsis always pointing in the direction of the Sun, allowing for consistent solar observation. Unlike eccentric orbits that experience precession in the Sun-Earth frame, the SEHO’s fixed orientation offers consistent power regeneration without large attitude maneuvers. The SEHO allows flexible change in orbit orientation by slightly changing the flyby geometry to place the satellite orbit in an identical resonance orbit but with a slightly smaller or larger Jacobi integral to rotate the orbit clockwise or counter-clockwise, respectively. 

The SEHO orbit has two oscillatory motions: Earth-Moon-plane planar oscillation and argument of periapsis oscillation per one sidereal year. The Earth-Moon-plane planar oscillation occurs as the Earth orbits the Sun, which changes the orbit's orientation. The argument of periapsis oscillation per one sidereal year occurs from the Moon’s inclination with respect to the ecliptic plane. 

\section{Optimal Sun-Earth Harmonic Orbit}

\begin{figure}
    \centering
    \includegraphics[page=4, width=1\linewidth]{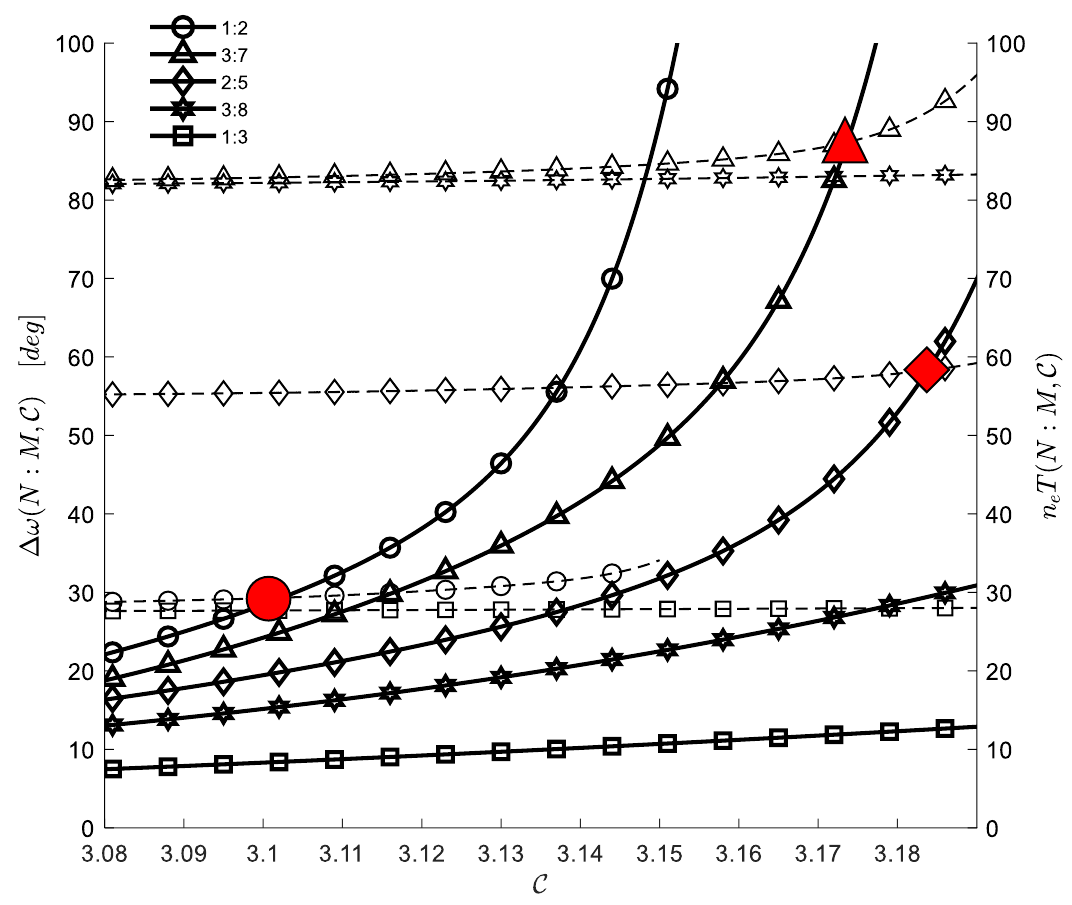}
    \caption{The argument of periapsis change versus Jacobi integral (solid line) with corresponding Earth’s true anomaly change per resonance orbit (dashed line)}
    \label{fig:optimalSEHOdEtermination}
\end{figure}

The resonance orbit search space was set as $C = \left[ {2.98,3.2} \right]$ for resonance orbit classifiers under N:M=1:2 with $N<4$. Resonance orbit with $N\geq4$ was ignored due to the large Earth-Moon-plane planar oscillation and large argument of periapsis change requirement per lunar flyby. Figure \ref{fig:optimalSEHOdEtermination} shows the change in $\Delta \omega \left( {N:M,C} \right)$ and ${n_e}T\left( {N:M,C} \right)$ for each resonance orbit with different Jacobi’s integral with a solid and dotted line, respectively. An increase in Jacobi’s integral results in an increase in the argument of periapsis change for each resonance orbit classifier. Also, as the resonance orbit classifier fraction ($N/M$) increases, the argument of periapsis variation increases, which is consistent for all resonance orbits. 
Note that the resonance orbits with the same N, but different M values have different ${n_e}T\left( {N:M,C} \right)$ at the same Jacobi integral due to the difference in the lunar flyby altitude.
Like $\Delta \omega \left( {N:M,C} \right)$, the orbital period of the resonance orbit also increases as the Jacobi integral increases and converges to a similar period as the Jacobi integral increase, which is visible in figure \ref{fig:optimalSEHOdEtermination}. 

\begin{figure}
    \centering
    \includegraphics[page=5, width=1\linewidth]{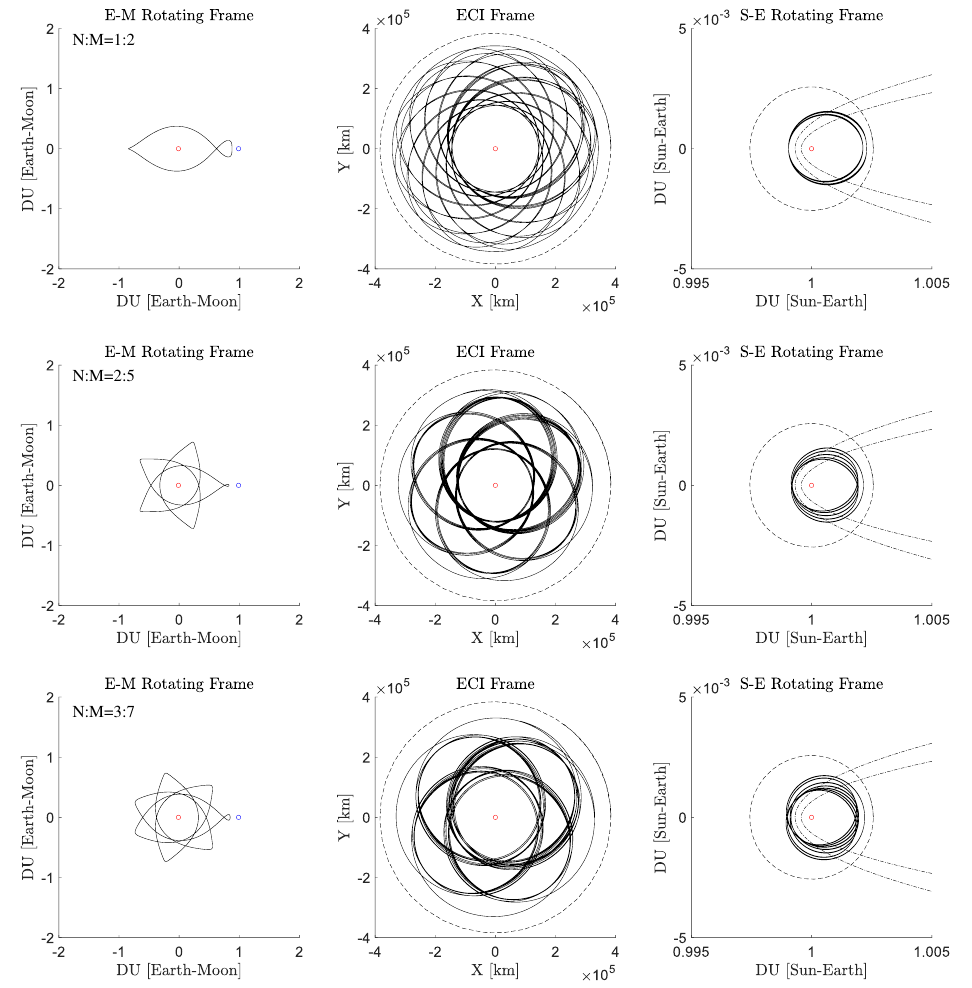}
    \caption{Optimal SEHO orbits in Earth-Moon rotating, Earth-Centered-Inertial (ECI), and Sun-Earth rotating frame.}
    \label{fig:optimaltrajectory}
\end{figure}

\begin{figure}
    \centering
    \includegraphics[page=9, width=1\linewidth]{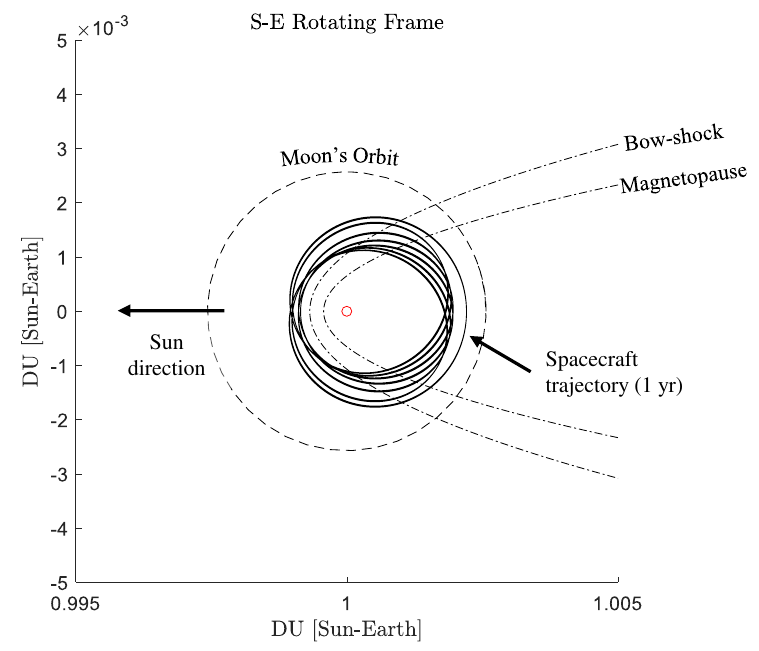}
    \caption{Optimal Sun-Earth Harmonic Orbit with Earth-Moon resonance orbit N:M=3:7.}
    \label{fig:optimaltrajectory37}
\end{figure}

\begin{figure}
    \centering
    \includegraphics[width=1\linewidth]{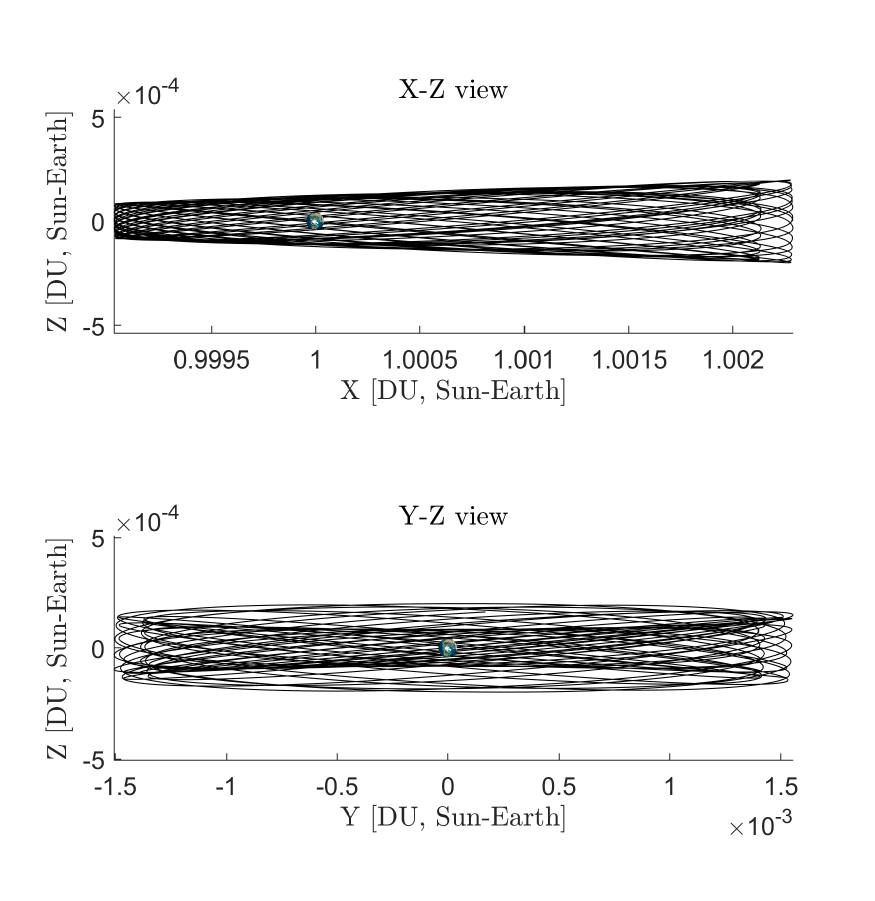}
    \caption{Oscillatory motion of the optimal Sun-Earth Harmonic Orbits }
    \label{fig:oscilatiory}
\end{figure}

The optimal SEHO is determined by selecting the resonance orbit with a specific Jacobi integral where $\Delta \omega \left( {N:M,C} \right)$ equals ${n_e}T\left( {N:M,C} \right)$, which is the point of intersection of the dashed (${n_e}T\left( {N:M,C} \right)$) and solid line ($\Delta \omega \left( {N:M,C} \right)$) in Figure \ref{fig:optimalSEHOdEtermination}, marked with red markers. From five resonance orbits, resonance orbit classifiers of 1:2, 3:7 and 2:5 have feasible SEHOs. The Jacobi integrals that produce the SEHO are C=3.109 for N:M=1:2, C=3.177 for N:M=3:7, and C=3.182 for N:M=3:7. The simulation parameters are tabulated in table \ref{tab:physicalparametrs}. The optimal SEHO is plotted in the ECI frame, Sun-Earth and Earth-Moon rotating frames in Figure \ref{fig:optimaltrajectory}. The trajectory was plotted for one sidereal year, which results in a 360-degree rotation of the argument of periapsis as shown in the ECI frame view. Figure \ref{fig:optimaltrajectory37} shows the N:M=3:7 optimal SEHO trajectory with labels of the bow-shock, magnetopause, Moon's orbit, and the Sun's direction with respect to the Sun-Earth rotating frame. The Earth-Moon-plane planar oscillation is visible in the Sun-Earth fixed frame, where the spacecraft performs lunar gravity assist to return to its original orbit. The resonance orbit with larger N experiences more prominent oscillation because the lunar flyby occurs less frequently. The oscillatory motions of the optimal SEHO, propagated for 1 year, are plotted in figure \ref{fig:oscilatiory}. The ideal region for performing the magnetospheric science missions, especially the magnetic reconnection, is about 12 $R_E$ and 25 $R_E$ from the Earth on the day and night side of the magnetosphere, which is in the range of the proposed orbits. The initial conditions of the optimal SEHO and their orbital periods are tabulated in table \ref{tab:ICs} with non-dimensional units of the Earth-Moon system.

\begin{table}
    \centering
    \caption{Physical parameters}
    \begin{tabular}{lll}
    \hline
         Parameter& Values & Units\\
         Mass parameter ($\mu$)& 0.0121536191408721 & - \\
         Gravitational acceleration ($g_0$)& 9.80665 & $m/s^2$\\
         Time unit (TU)& 377498.438 & $s$ \\
         Distance unit (DU)& 384400 & $km$\\
         Speed unit (VU)& 1.01828235817878 & $m/s$ \\
         Mean motion of Earth ($n_e$)& 1.99096871e-7 & $rad/s$\\
        \hline
    \end{tabular}
    \label{tab:physicalparametrs}
\end{table}

\begin{table}
    \centering
    \begin{tabular}{ccccc}
    \hline
        N:M & $x$ & $v_y$ & $C_{opt}$ & $T(N:M,C)$\\
         1:2& 0.8782432288&  -0.3344655870&   3.100109045&  6.799697050\\
         3:7& 0.8475817753 & -0.1210038504 &  3.175072751& 20.370740880\\
         2:5& 0.8288107874 & -0.0565351140&   3.185890533& 13.592628156\\
    \hline
    \end{tabular}
    \caption{Caption}
    \label{tab:ICs}
\end{table}

\section{Application of the Sun-Earth Harmonic Orbit}

This section analyzes the stability and quantifies the orbit-orientation reconfiguration cost for each optimal SEHO presented in the previous section. The bi-circular restricted four-body problem (BRFBP) is incorporated to include the effect of the Sun on the candidate SEHO orbits. The potential function of the BRFBP is as follows: 
\begin{equation}
{\Omega _4} = {\Omega _3} + \frac{{{\mu _{sun}}}}{{{r_3}}} - \frac{{{\mu _{sun}}}}{{{{\left\| {\bf{L}} \right\|}^3}}}\left( {{L_x}x + {L_y}y + {L_z}z} \right)
\end{equation}
Where $\Omega_3$ is from Eq. \ref{eq:omega3}, and $\mu_s(=m_s/(m_1+m_2))$ denotes the mass gravitational parameter of the Sun in Earth-Moon canonical units.

${\bf{L}}\left( { = \left[ {{L_x},{L_y},{L_z}} \right] = \left[ {L \cdot \cos \left( {{\theta _S}} \right),L \cdot \sin \left( {{\theta _S}} \right),0} \right]} \right)$
denotes the Sun’s position vector from the Earth-Moon barycenter, where $\theta_s(=\theta_0-\omega_s t)$ and  $\omega_s(=-\sqrt{(1+\mu_s/\vert L \vert ^3)})=-0.9253rad/s$ denote the angular offset and speed of the Sun measured from the Earth-Moon system’s x-axis. The distance from two barycenters ($L$) was set as 392.1DU.

\subsection{Stability of the Optimal Sun-Earth Harmonic Orbits}

The stability of the resonance orbit is determined by utilizing the Monodromy matrix. The Monodromy matrix (${\cal M}$) is the state transition matrix integrated along the resonance orbit’s trajectory for one orbital period as follows.
\begin{equation}
{\cal M} = \Phi \left( {T\left( {N:M,C} \right),{t_0}} \right)
\end{equation}
The Monodromy matrix is integrated along the trajectory with the following differential equation. 
\begin{equation}
\dot \Phi \left( {t,{t_0}} \right) = A\left( t \right) \cdot \Phi \left( {t,{t_0}} \right)
\end{equation}
$A(t)$ is defined by using the potential function from eq. \ref{eq:omega3} as follows. 
\begin{equation}
\begin{array}{*{20}{c}}
{A\left( t \right) = \left[ {\begin{array}{*{20}{c}}
{{0_{3 \times 3}}}&{{I_{3 \times 3}}}\\
{\cal U}&\zeta 
\end{array}} \right]}
\end{array}
\end{equation}
where
\begin{equation}
\begin{array}{*{20}{c}}
{\zeta  = \left[ {\begin{array}{*{20}{c}}
0&2&0\\
{ - 2}&0&0\\
0&0&0
\end{array}} \right]}
\end{array}
\end{equation}

\begin{equation}
    {{\cal U} = \left[ {\begin{array}{*{20}{c}}
{{\Omega _{xx}}}&{{\Omega _{xy}}}&{{\Omega _{xz}}}\\
{{\Omega _{yz}}}&{{\Omega _{yy}}}&{{\Omega _{yz}}}\\
{{\Omega _{zx}}}&{{\Omega _{zy}}}&{{\Omega _{zz}}}
\end{array}} \right]}
\end{equation}
The second-order partial derivatives of $\Omega_3$(Eq. \ref{eq:omega3} for the ${\cal U}$ matrix are presented in Appendix. 

\begin{figure}
    \centering
    \includegraphics[page=6, width=1\linewidth]{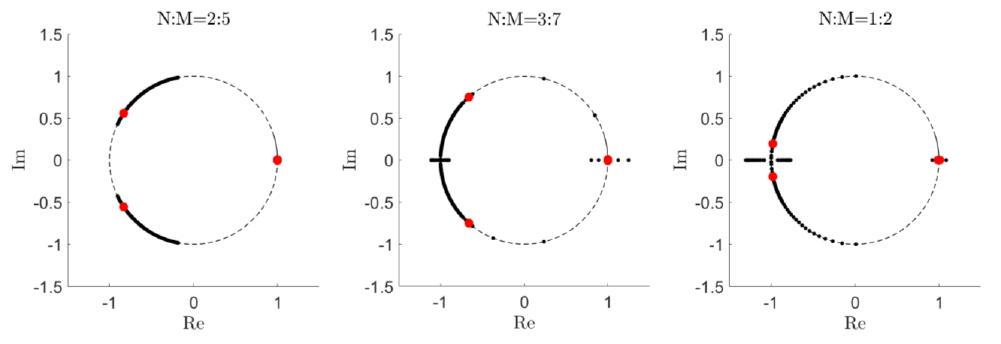}
    \caption{Stability analysis of the 2:5, 3:7, and 1:2 resonance orbits. Red dots represent the stability roots for the optimal Sun-Earth Harmonic Orbits for each N:M}
    \label{fig:stability}
\end{figure}

The ordinary differential equation is propagated along the resonance orbit’s trajectory with the identity matrix as the initial condition. The stability of an orbit, especially periodic and resonance, is determined based on Broucke’s Stability Diagram \cite{broucke1969stability}. Following the notions from \cite{campbell1999bifurcations},\cite{grebow2010trajectory},\cite{bolliger2019cislunar}, the eigenvalue-eigenvector analysis was performed on the Monodromy matrices for all SEHO resonance orbit classifiers. The results are shown in Figure \ref{fig:stability}. As the trajectory is symmetric about the x-axis of the rotating frame, the roots of the monodromy matrix is along the unit circle of the root-locus diagram. Simulation results show that the optimal SEHO orbits’ Earth-Moon resonance orbits are in region VI of the Broucke diagram, representing an even-semi-instability. Though the orbit is in the region of instability, the even-semi-instability offers unstable and stable manifolds from the resonance orbit, which can be studied further for station-keeping analysis.

\subsection{Orbit-Orientation Reconfiguration and Station Keeping of Optimal Sun-Earth Harmonic Orbit}

\begin{figure}
    \centering
    \includegraphics[page=7, width=1\linewidth]{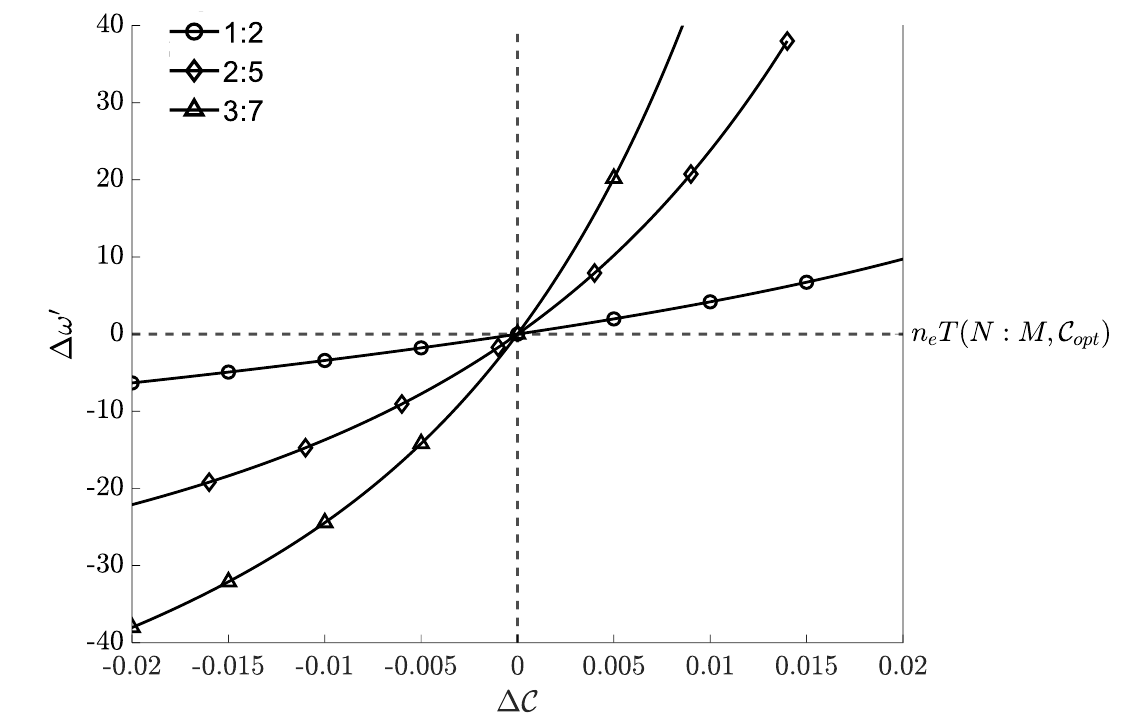}
    \caption{Orbit shifting with Jacobi integral change in Sun-Earth rotating frame}
    \label{fig:stationkeep}
\end{figure}

\begin{figure}
    \centering
    \includegraphics[page=8, width=0.65\linewidth]{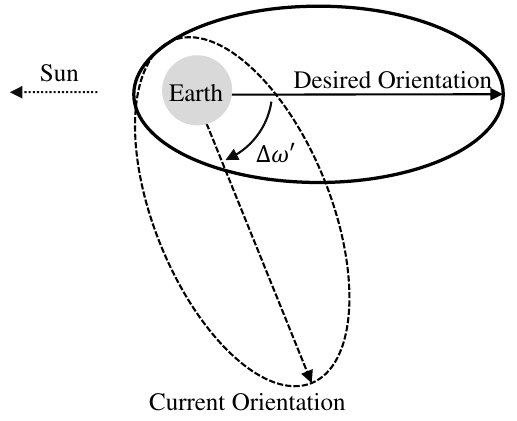}
    \caption{Definition of $\Delta \omega '$}
    \label{fig:definitionomega}
\end{figure}

The spacecraft's orbit experiences a gradual shift due to the fourth-body perturbation from the Sun in the BRFBP dynamics. Therefore, station-keeping maneuvers are necessary to keep the orbit orientation stationary in the rotating frame. Instead of using the manifold dynamics determined by the stability index, we propose an alternative method: to slightly adjust the Jacobi integral of the resonance orbit to generate an orientation change, thereby compensating for the fourth-body perturbation.
Figure \ref{fig:stationkeep} quantifies the amount of orbit orientation shift measured in the Sun-Earth rotating frame ($\Delta \omega '$) by varying the Jacobi integral ($\Delta C$) from the optimal Jacobi integral ($C_{opt}$) for each resonance orbit classifier. The definition of the new angle $\Delta \omega '$ is shown in Figure \ref{fig:definitionomega}.

At $\Delta \omega ' = 0$ and $\Delta C = 0$, the orbit represents its optimal configuration ($\Delta \omega = {n_e}T\left( {N,{C_{opt}}} \right)$). The orbit-orientation reconfiguration method offers several advantages over relying on the resonance orbit manifold for determining station-keeping maneuvers. The stable manifolds in the rotating frames are propagated backward in time, meaning that the terminal time for station-keeping maneuvers is predetermined. In contrast, the orbit-orientation reconfiguration method allows for more flexibility, as it quantifies $\Delta C$ to generate the desired $\Delta \omega '$, enabling easier and more adaptable determination of station-keeping maneuvers.
For example, a satellite in a 2:5 resonance orbit that requires a reconfiguration of $\Delta \omega ' = 10^\circ $ could either adjust the Jacobi integral by -0.005 and perform a single lunar flyby, or by -0.0015 and perform three lunar flybys before returning to ${C_{opt}}$. The trajectory optimization problem with resonance orbits is further explored in \cite{campagnola2012flybys} and \cite{lee2024low}.

\section{Conclusion}
This paper introduces the concept of the Sun-Earth harmonic orbit, which leverages continuous gravitational assists from the Moon in Earth-Moon resonance orbits to maintain its orientation relative to the Sun. The orbit's orientation is adjusted by the rotation of the spacecraft's argument of periapsis within the equatorial plane.
We successfully computed three internal resonance orbits ($N<M$) that exhibit the characteristics of the Sun-Earth harmonic orbit. The optimal Earth-Moon harmonic orbits were identified using resonance ratios of 1:2, 3:7, and 2:5. Through stability analysis, we determined that these optimal resonance orbits lie within the unstable region of Broucke's stability diagram.
To address this instability, we proposed a station-keeping strategy that relies on gravity-assist maneuvers instead of conventional propulsion, thereby conserving station-keeping propellant mass.

\bibliography{sample}

\end{document}